# Anisotropy of the irreversibility field for Zr-doped (Y,Gd)$Ba_2Cu_3O_{7-x}$ thin films up to 45T

C. Tarantini[1], J. Jaroszynski[1], F. Kametani[1], Y. L. Zuev[2,3], A. Gurevich[1,4], Y. Chen[5],

V. Selvamanickam[6], D. C. Larbalestier[1], D. K. Christen[2]

[1]National High Magnetic Field Laboratory, Florida State University, Tallahassee, Florida 32310, USA

[2]Oak Ridge National Laboratory, Oak Ridge, Tennessee 37831, USA

[3]University of Tennessee, Knoxville, Tennessee 37996, USA

[4] Department of Physics, Old Dominion University, Norfolk, Virginia 23529, USA

[5] SuperPower Inc., Schenectady, New York 12304, USA

[6] Department of Mechanical Engineering and the Texas Center for Superconductivity,

University of Houston, Houston, Texas 77204, USA

The anisotropic irreversibility field $B_{Irr}$ of two $YBa_2Cu_3O_{7-x}$ thin films doped with additional rare earth (RE)=(Gd,Y) and Zr and containing strong correlated pins (splayed $BaZrO_3$ nanorods, and $RE_2O_3$ nanoprecipitates), has been measured over a very broad range up to 45T at temperatures 56 K<T <$T_c$. We found that the experimental angular dependence of $B_{Irr}$ (θ) does not follow the mass anisotropy scaling $B_{Irr}(\theta) = B_{Irr}(0)(\cos^2\theta + \gamma^{-2}\sin^2\theta)^{-1/2}$, where $\gamma = (m_c/m_{ab})^{1/2}$ = 5-6 for the RE-doped $YBa_2Cu_3O_{7-x}$ (REBCO) crystals, $m_{ab}$ and $m_c$ are the effective masses along the *ab* plane and the *c*-axis, respectively, and $\theta$ is the angle between **B** and the *c*-axis. For **B** parallel to the *ab*-planes and to the *c*-axis correlated pinning strongly enhances $B_{Irr}$, while at intermediate angles, $B_{Irr}(\theta)$ follows the scaling behavior $B_{Irr}(\theta) \propto (\cos^2\theta + \gamma_{RP}^{-2}\sin^2\theta)^{-1/2}$ with the effective anisotropy factor $\gamma_{RP} \approx 3$ significantly smaller than the mass anisotropy would suggest. In spite of the strong effects of *c*-axis $BaZrO_3$ nanorods, we found even greater enhancements of $B_{Irr}$ for fields along the *ab*-planes than for fields parallel to the *c*-axis, as well as different temperature dependences of the correlated pinning contributions to $B_{Irr}$ for *B//ab* and *B//c*. Our results show that the dense and strong pins, which can now be incorporated into REBCO thin films in a controlled way, exert major and diverse effects on the measured vortex pinning anisotropy and the irreversibility field over wide ranges of *B* and *T*. In particular, we show that the relative contribution of correlated pinning to $B_{Irr}$ for *B//c* increases as the temperature increases due to the suppression of thermal fluctuations of vortices by splayed distribution of $BaZrO_3$ nanorods.

## I. INTRODUCTION

In recent years the properties of $YBa_2Cu_3O_{7-x}$ (YBCO), or more generally any rare-earth- (RE-) doped $YBa_2Cu_3O_{7-x}$ (REBCO) thin films, have been progressively improved by mitigating the grain boundary limitation of the critical current density $J_c$ and by increasing the vortex pinning by incorporating non-superconducting nanoparticles or nanorods. Pure YBCO exhibits an upper critical field anisotropy $\gamma = B_{c2}^{(ab)}/B_{c2}^{(c)} = (m_c/m_{ab})^{1/2}$ of 5-6 , where $m_c/m_{ab}$ is the mass anisotropy ratio ~25-36. The substitution of RE on the Y site (or the enrichment of Y or RE) produces $RE_2O_3$ nanoparticle[1,2,3] pinning centers while the addition of Zr induces self-assembled $BaZrO_3$ (BZO) nanorods able to strongly increase $J_c$ for B//*c*.[1,2,4] Blatter's scaling approach[5] for weak isotropic random pinning (RP) has often been used[6,2] to separate the RP contribution to the angular dependence of $J_c$ determined by the mass anisotropy from the contributions of correlated pins which are most pronounced for the magnetic fields parallel to the *ab*-planes or to the *c*-axis. The scaling approach[5] enables one to calculate the angular dependence of $J_c(H,\theta)$ by rescaling the field dependence of $J_c(H)$ for $B//c$ according to $B \rightarrow B\varepsilon_\theta$ where $\varepsilon_\theta = (\cos^2\theta + \gamma^{-2}\sin^2\theta)^{1/2}$ . Here $\gamma = (m_c/m_{ab})^{1/2}$ = 5-6 for the REBCO, $m_{ab}$ and $m_c$ are the effective masses along the *ab* plane and the *c*-axis, respectively, and $\theta$ is the angle between **B** and the *c*-axis. This approach works well for clean YBCO films with weak correlated pinning where $\gamma$~5 is actually found.[7] However, for films with high density of *c*-axis aligned nanorods (like BZO) , the angular dependence of $J_c$ for the fields tilted away from the symmetry axis was found to follow the scaling behavior with $\varepsilon_\theta = (\cos^2\theta + \gamma_{eff}^{-2}\sin^2\theta)^{1/2}$ and the anisotropy parameter $\gamma_{eff}$ smaller than $\gamma$.[2,6] Such reduced anisotropy was ascribed by the authors to the effect of strong RP after the effect of correlated pinning has been separated from the angular dependence of $J_c$ .

The complex effects of strong random and correlated pinning on the angular dependence of the irreversibility field $B_{Irr}$ and particularly, the question if the contributions of random and correlated pinning on $B_{Irr}(\theta)$ can be identified from the observed angular dependence $B_{Irr}(\theta)$ are not well understood. This is not an easy task because the relative weights of random and correlated contributions can change as the temperature decreases and the role of thermal fluctuations of vortices diminishes. In turn, the observation of these changes also requires measurements of $B_{Irr}(\theta)$ at very high magnetic fields, which was the main goal of this work. Because the irreversibility field $B_{Irr}$ at low temperatures becomes much larger than the matching field, $B_\phi$, of the artificially introduced defects (seldom greater than 3-5 T), $B_{Irr}$ might be less affected by the correlated pinning than $J_c$, making it easier to reveal the effect of strong RP on $B_{Irr}$. This is the central conjecture of this paper, in which we present the angular dependence of $B_{Irr}$ performed in very high dc fields up to 45 T. Access to such high fields allows us to conduct $B_{Irr}$ measurements down to temperatures of 56 K where thermal fluctuations weaken and the pinning behavior starts to change. We show that such high fields allow us to see that multiple different effects are present in $B_{Irr}$ and to check how the effective anisotropy of $B_{Irr}$ differs from the mass anisotropy. In particular, we found that the angular dependence $B_{Irr}(\theta)$ for the intermediate field orientations not too close to the symmetry axis approximately follows the scaling behavior $B_{Irr}(\theta)=B_{Irr}(0)(\cos^2\theta+\gamma_{RP}^{-2}\sin^2\theta)^{-1/2}$ with the effective anisotropy parameter $\gamma_{RP}\approx 3$, significantly smaller than the mass anisotropy $\gamma = 5\text{-}6$ for REBCO. This behavior of $B_{Irr}(0)$ appears qualitatively similar to the angular dependence of $J_c(\theta)$ discussed above, although the parameters $\gamma_{eff}$ and $\gamma_{RP}$ may be different. Moreover, we were also able to identify and measure the strength of the correlated pinning contribution to $B_{Irr}$ that appears to be particularly effective for $B//ab$ and for $B//c$ at temperatures close to the critical temperature, $T_c$. We also investigated the behavior of the electrical resistance $R(B)\propto(B-B_{Irr})^s$ near the depinning transition by studying the critical exponents $s$ and the way that the pinning mechanisms change with temperature. Finally we compare our results with those previously reported in underdoped single crystals,[8] optimally doped thin films[9] and thin films of REBCO containing additional pinning centers.[10]

## II. EXPERIMENTAL RESULTS AND ANALYSIS

The samples studied in this work are ~ 1-μm-thick and ~0.5-μm-thick (Gd,Y)BCO thin films, both made with additional Zr and rare earth (RE). The superconducting films were deposited by metal organic chemical vapor deposition (MOCVD) on Ion Beam Assisted Deposition (IBAD)-MgO template, which has a multilayer structure of $LaMnO_3$/homoepitaxial-MgO/IBAD-MgO/$Y_2O_3$/$Al_2O_3$/Hastelloy tape. The growth process of such samples has been described previously.[3] Here, compositions of the liquid precursors delivered for MOCVD deposition of the two samples were $Zr_{0.1}Gd_{0.66}Y_{0.66}Ba_2Cu_3$ and $Zr_{0.05}Gd_{0.65}Y_{0.65}Ba_2Cu_3$ identified in the following as 10%Zr-REBCO (1 μm) and 5%Zr-REBCO (0.5 μm). The films have a critical temperature of 90.2 and 91.0 K, respectively. The microstructure of 5%Zr-REBCO was examined in a JEOL JEM2011 transmission electron microscope (TEM). As was previously reported[1-4], the Zr and RE additions greatly enhance vortex pinning by introducing nanoprecipitates and nanorods in the REBCO matrix as shown in the transverse cross-sectional TEM images of Fig.1. Figure 1(a) shows the entire thickness of the 5%Zr-REBCO film in which three different pinning defects can be seen. The contrasts of vertical dark rods, black dots and horizontal white lines represent the BZO nanorods, $RE_2O_3$ nano-precipitates and stacking faults, respectively, all of which are distributed quite uniformly in the REBCO matrix. The BZO nanorods form in a splayed manner,[4] but with a statistically averaged angle between the direction of nanorods and the REBCO *ab*-planes of approximately 80° rather than being 90°, as indicated in the high resolution TEM (HREM) image of Fig. 1(b). In addition to the ~5-nm-diameter BZO nanorods, Fig. 1(b) also shows a high density of stacking faults with an average spacing of ~10 nm and $RE_2O_3$ nanoprecipitates that are ~6 nm thick and ~7 nm in diameter. HREM imaging of the interface between the REBCO and the buffer revealed that the REBCO *ab*-planes are not parallel to, but are tilted ~3° from the interface and thus the film surface [Fig. 1(c)]. TEM imaging previously performed on the 10%Zr-REBCO sample showed a similar microstructure, although with a larger density of $RE_2O_3$ precipitates arranged in a plane slightly tilted from the *ab* planes.[11]

The angular dependent transport measurements were performed in a Quantum Design 16 T Physical Property Measurement System (PPMS) and in a 31 T resistive and a 45 T hybrid dc magnets at the National High Magnetic Field Laboratory (NHMFL). Electrical resistivity measurements were made using a four-wire, ac resistance bridge at a current density of <100 A/cm$^2$, with the magnetic field applied perpendicular to the current in a maximum Lorentz force configuration. Measurements were made in ramped field, constant-temperature mode, with the irreversibility field defined at 1% of the normal state resistance just above the superconducting transition. The angular dependencies of $B_{Irr}(\theta)$ are shown in Fig. 2 as filled (PPMS data) and empty (high dc field data) symbols where $\theta$ is the angle between the field and the *c*-axis (the field alignment with the main crystallographic axes was determined by fits of the RP contribution to $B_{Irr}(\theta)$ , but not relatively to the tape axes, as explained later). All the data show marked peaks close to 90° and 270°, where the field is aligned with the *ab* plane and an intermediate, almost flat region which has a small peak at about 190°. This almost flat region is ~80°-90° wide at higher temperatures, but narrows to ~40° with decreasing temperature.

The angular dependency of the irreversibility field $B_{Irr}(\theta)$ is strongly affected by the sample nanostructure, as was found also for $J_c(\theta)$.[4] However, at high magnetic fields ($B \gg B_\phi$, where $B_\phi \leq 5$ T), the contribution of the *c*-axis-correlated pinning effects becomes less pronounced, and therefore the correlation of angular dependence with the REBCO mass anisotropy in $B_{Irr}$ should become more explicit than in the critical current measurements. In order to analyze $B_{Irr}(\theta)$ we have to take into account different contributions that may affect its behavior. The irreversibility field in the presence of weak RP roughly scales like the vortex melting field which mainly depends on the mass anisotropy.[5]

Thus, to separate the RP contribution $B_{Irr,RP}$ in the observed $B_{Irr}(\theta) = B_{Irr,\,cor} + B_{Irr,RP}$ we use a scaling expression for $B_{Irr,RP}(\theta)$ similar to the one for the melting field,[5] but assuming that the effective anisotropy parameter $\gamma_{RP}{}^{2}$ might differ from $\gamma^{2} \approx 25-36$

$$B_{Irr,RP}(\theta) = \frac{B_{Irr}(0)}{\sqrt{\cos^{2}(\theta) + \gamma_{RP}{}^{-2}\sin^{2}(\theta)}} = \frac{B_{Max}}{\sqrt{\sin^{2}(\theta) + \gamma_{RP}{}^{2}\cos^{2}(\theta)}} \quad (1)$$

with $B_{Max} = \gamma_{RP} B_{Irr}(0)$. The question to be addressed is then how the correlated pinning defect nanostructure of the REBCO thin films would affect the angular dependence of $B_{Irr}$. Columnar BZO nanorods produce strong correlated pinning for the field nearly parallel to the *c*-axis direction affecting the region around 180° in Fig. 2. Planar second-phase defects, stacking faults and the intrinsic pinning by the REBCO layered structure can give additional pinning contributions for fields parallel to the *ab* planes that are slightly misoriented from the direction of the surface because of the ~3° tilted REBCO planes [Fig 1(c)]. The data in Fig. 2 clearly show all these contributions.

To deconvolute different correlated pinning contributions causing deviations of the observed $B_{Irr}(\theta)$ from the $B_{Irr,RP}(\theta)$ scaling, we rewrite Eq. (1) as follows

$$\frac{1}{B_{Irr,RP}^{2}(\theta)} = \frac{1}{B_{Max}^{2}} + \frac{\gamma_{RP}^{2} - 1}{B_{Max}^{2}}\cos^{2}(\theta) \quad (2)$$

This form is convenient because correlated pinning contributions to $B_{Irr}(\theta)$ on top of the RP component would manifest themselves as deviations of $1/B_{Irr}^{2}(\theta)$ plotted as a function of $\cos^{2}\theta$ from a straight line. Figure 3 shows an example of this approach using the replotted data from Fig. 2 (10%Zr-REBCO at 72.6 K). At intermediate orientations between *B*//*c* and *B*//*ab* a good fit to Eq.2 can be clearly seen, while there are deviations for $\cos^{2}(\theta)$ close to 0 and 1 due to the correlated pinning contributions. These RP contributions to $B_{Irr}(\theta)$ are shown in Fig. 2 as continuous lines for which the parameters $B_{Max}$ and $\gamma_{RP}$ have been estimated from the best fits to Eq. (2) for every data series. This analysis yields for both samples $\gamma_{RP}$~ 3, which is smaller than $\gamma$ = 5-6 for REBCO single crystals. Possible reasons for the reduced anisotropy parameter $\gamma_{RP}$ will be discussed below. Because of the different temperature and field ranges over which the two samples were characterized, in the following we mainly focus on the *c*-axis peak of 5%Zr-REBCO (measured down to 56K) and on the *ab* peak of 10%Zr-REBCO, which was better characterized in the 45T magnet.

Close inspection shows that the experimental data in Fig. 2 deviate from the continuous line in an asymmetric way in the regions indicated by the arrows in the inset of Fig. 2(a) for 10%Zr-REBCO, whereas there is no appreciable asymmetry for 5%Zr-REBCO [see 77K data in Fig. 2(b)]. Such asymmetry indicates the presence of correlated pinning not perfectly aligned with the *ab* planes of 10%Zr-REBCO. Subtracting the RP contribution to the experimental data, we extract the correlated pinning contributions $\Delta B_{Irr}(\theta) = B_{Irr}(\theta) - B_{Irr,RP}(\theta)$ that are shown in Fig. 4. The most pronounced effect of correlated pinning is that $B_{Irr}$ along the *ab* planes is strongly enhanced from 20 to 45 T at 78 K, more than doubling the expected RP component of $B_{Irr}$. The inset of Fig. 4(a) shows that the shape of the *ab*-peak for the 10%Zr-REBCO sample becomes cusp-like on decreasing the temperature from 84K to 78K. This effect is emphasized in the inset of Fig. 5 where the $\Delta B_{Irr}(\theta)$ data are plotted on a logarithmic scale. Here all the curves for T>78K follow the same behavior, but $B_{Irr}$(78K) clearly has a double structure consisting of wide and narrow peaks. The inset of Fig. 4(a) (continuous lines) shows that the high temperature data can be described by a single Lorentzian peak [full width half maximum (FWHM)~12°], whereas the 78K data exhibit a superposition of two Lorentzian peaks (the light blue curves). The sharper (FWHM~1.6°) and the most intense peak is aligned with the *ab* plane, while the wider (FWHM~12°) and weaker one is slightly shifted to the low-angle side, as already noted in Fig.2a. Normalizing the *ab*-peak-correlated pinning to the RP contribution (Fig. 5, main panel), we observe that the wider peak becomes stronger at high temperatures, while the narrower one emerges only at low temperatures, clearly suggesting different mechanisms that affect $B_{Irr}$. The narrower peak seems not to be detected below 78K but this may be because $B_{Irr}$ close to *ab* is beyond the 45T limit of our measurements.

For fields nearly parallel to the *c*-axis, the inset of Fig. 4(b) shows the peaks of $\Delta B_{Irr}$ for the 5%Zr-REBCO sample. The peak magnitude increases with decreasing temperature and its position, which at higher temperatures is at over 190°, moves closer to 180° at lower temperatures. However, the $\Delta B_{Irr}$ data normalized to the RP contribution indicate that the relative contribution of the correlated pinning decreases at lower temperature (Fig. 6), a result that is different from what we found for the *ab*-plane peak. In the 10%Zr-REBCO, the higher Zirconium content does not appear to have a positive effect on the *c*-axis pinning contribution to $B_{Irr}$. The reason can be related to the fact that, in this relatively high-temperature regime, the irreversibility field is reduced due to the slightly suppressed critical temperature of 10%Zr-REBCO.

## III. DISCUSSION

### A. Random pinning

The effective anisotropy parameter $\gamma_{RP} \approx 3$ experimentally found for the intermediate range of field orientations far from the main crystallographic axes, is significantly lower than the mass anisotropy parameter $\gamma \approx 5-6$. This reduced value of $\gamma_{RP}$ indicates the effect of RP on $B_{Irr}$. Here several mechanisms affecting the $B_{Irr}(\theta)$ can be identified. For weak pinning by randomly distributed point defects, $B_{Irr}$ is defined by the condition $U(B_{Irr},T) \sim k_BT$ where $U$ is the activation barrier which determines the electric field $E \propto J\rho_n \exp[-U/k_BT]$ due to thermally activated hopping of vortices. In the scaling approach, $U(B_{Irr},T)$ depends only on the product $B\varepsilon_\theta$ where $\varepsilon_\theta=(\cos^2\theta + \gamma^{-2}\sin^2\theta)^{1/2}$ and $\theta$ is the angle between the $c$-axis and **B**. This results in Eq. (1) in which the magnitude $B_{Max}$ depends on the pinning strength.[5]

The actual structure of randomly distributed pinning nanoprecipitates can change the scaling angular dependence of $B_{Irr,RP}$ due to (1) shape anisotropy of nanoprecipitates, (2) anisotropy of core and magnetic interaction of a vortex with nanoprecuipitates, (3) different mean spacing between the precipitates in the $ab$ planes and along the $c$-axis. For instance, the magnetic interaction force $f_p$ of a vortex with a round precipitate is anisotropic so that $f_p(B//c)\approx\gamma f_p(B//ab)$, reflecting the decrease of screening currents in a vortex upon rotating toward the $ab$ plane. In turn, the reduction of both $f_p$ and the line tension of a vortex parallel to the $ab$ planes have opposite effects on $J_c$ [6] and can also result in a weaker angular dependence of $B_{Irr}(\theta)$ as compared to what is expected from the effective mass anisotropy only. Moreover, oblate nanoprecipitates can enhance pinning of vortices parallel to the $ab$ plane, making $f_p(B//ab)$ larger than $f_p(B//c)$ if the aspect ratio of precipitates is greater than $\gamma$. Thus the observed reduction of $\gamma_{RP}$ can result from multiple conflicting mechanisms of pinning of vortices by nanoprecipitates.

For the dense and strong pinning nanostructures of the samples studied here, pinning defects can also change the global electronic properties of the conductor, particularly the effective anisotropy parameter $\gamma_{eff}$ due to strains around oxide nanoprecipitates, buckling of the $ab$ planes, crystal mosaicity, and low-angle grain boundaries caused by dense pinning nanostructure result in local variations of the orientation of the $c$-axis with respect to the applied field. Also the local mass anisotropy ratio in the cuprates can change if defects cause local nonstoichiometry variations increasing $\gamma(\mathbf{r})$ in underdoped regions.

For instance, in the presence of a distribution of local orientations of the *c*-axis on scales greater than the coherence length, the global irreversibility field $B_{Irr}$ is given by averaging the local relation Eq. (1)

$$B_{Irr} = B_{max}\int_0^{\pi}\frac{F(\phi)d\phi}{\sqrt{sin^2\phi+\gamma^2 cos^2\phi}} \qquad (3)$$

where F(ϕ) is the distribution function of the angles ϕ(**r**) between the local orientation of the *c*-axis and the direction of the applied field, F(ϕ) is normalized by the condition $\int_0^{\pi} F(\phi)d\phi = 1$. The global irreversibility field $B_{Irr}$ $(\theta)$ can be for instance calculated for a narrow step-wise distribution F(ϕ) = $1/2\theta_0$ for $|\phi-\theta|<\theta_0$ and F(ϕ)=0 for $|\phi-\theta|>\theta_0$ where $\theta_0$ quantifies the spread of the local *c*-axis orientations around the angle $\theta$ between **B** and the mean orientation of the *c*-axis. For $\theta_0<<1$ the parameter $\gamma_{eff}$, defined as $B_{Irr}(\pi/2)/ B_{Irr}(0)$, decreases as $\theta_0$ increases as shown in Fig. 7. However in this example the reduction of $\gamma$ from 5 to the observed ≈3 corresponds to $\theta_0\approx30^o$. Such a large angular spread of misorientation angles suggests that this cannot be the only mechanism reducing $\gamma$. Another characteristic feature of samples with a large content of nanoparticles is that the REBCO lattice is buckled around pinning nanoparticles modulating the distance between the *ab* planes. In turn, a local reduction of the *ab* distances due to strains might reduce the suppression of the order parameter between the planes and decrease the effective mass anisotropy.

### B. Correlated pinning

The *ab* peak in the 5%Zr-REBCO appears to have a single Lorenzian structure [Fig. 4(b)] in the temperature and field range of our measurements even if, because the magnitude of $B_{Irr}$ becomes higher than 45 T, the full angular dependence of the *ab* peak was not measured. This single-peak structure is consistent with our TEM images (Fig. 1), showing a high density of stacking fault unevenly spaced by distances $d_s$~10 nm along the *c*-axis but no significant arrangement of the nanoprecipitates along the *ab* direction. Because of the short *c*-axis coherence length $\xi_c$~0.3-0.5 nm in REBCO, stacking faults locally depress superconductivity, resulting in strong core and magnetic pinning of parallel vortices. For fields lower than the matching field, $B<B_s\sim\phi_0/\gamma d_s^2\approx 4$ T at $\gamma=5$, vortex cores are mostly trapped by the stacking faults, so misalignment of B with the stacking faults causes a cusp-like drop of the pinning force,[5] which manifests itself in the *ab* peak in $B_{Irr}$. For $B>B_s$ vortices also appear between the stacking faults which cut off currents circulating around vortices because $d_s$~10 nm is much smaller than the London penetration depth $\lambda$(77K)≈400 nm along the *c*-axis. This results in strong magnetic pinning of vortices if the Josephson critical current densities across the stacking faults are smaller than $J_s\sim J_d\xi_c/d_s$ where $J_d$ is the depairing current density in the *ab* plane.[12] In this case stacking faults effectively subdivide REBCO films into a stack of weakly coupled layers of different thickness distributed around ~ 10 nm<< $\lambda$(77K) which would result in a cusp-like peak in $J_c$, manifesting itself in the *ab* peak in $B_{Irr}$. Such strong cusps in $J_c$ were indeed observed even for isotropic NbTi films in which the *ab* peak in $J_c$ not only reached ~ (0.1-0.2)$J_d$ for a thin film with d = $\lambda/2$ but remained quite noticeable even for thicker films with d=2$\lambda$.[13] These results suggest that the long-range magnetic pinning of vortices by the sample surfaces could be essential for our REBCO films given that the thicknesses of 5%Zr-REBCO film (0.5 μm) and 10%Zr-REBCO film (1 μm) are both comparable to $\lambda$(77K)~0.4 μm, thus, the thickness effect could contribute to the *ab* peaks in $B_{Irr}$. As is clear from Figs. 2(b) and 4(b), in this case the correlated pinning may enhance $B_{Irr}$ parallel to the *ab* planes by as much as 50-100%, depending on temperature.

In the case of 10%Zr-REBCO, the double structure of the *ab* peak in $B_{Irr}$ observed in Figs. 4(a) and 5 suggests that the peak may be determined by a superposition of different pinning mechanisms that can be roughly described by two Lorentz distributions in $\Delta B_{Irr}(\theta)$. The wider Lorentz peak in $\Delta B_{Irr}(\theta)$ exists at all temperatures and is shifted by 2-3° with respect to the maximum of the RP contribution. Although several possible mechanisms (such as pinning by stacking faults and by film surface) may contribute to the broad peak in $\Delta B_{Irr}(\theta)$, the 2-3° shift suggests also the influence of pinning by $RE_2O_3$ precipitates that in this sample grow in the *ab* planes but stack in a slightly tilted staircase manner, as observed in these same materials by Zhang *et al.*[11] and in similar materials by Holesinger *et al.*[14]

A distinctive feature of the 10%Zr-REBCO film is the emergence of an intense narrow *ab* peak in $\Delta B_{Irr}(\theta)$ as the temperature is decreased. A similar strong increase of $B_{Irr}$ for B along the *ab* planes was first observed in underdoped YBCO crystals[8] and later also in optimally doped YBCO films.[9] In both cases the authors found that there is a characteristic temperature $T_{cf}$ below which $B_{Irr}$ increases very steeply with decreasing temperature for the field parallel to the *ab* planes. This characteristic temperature was estimated to be about 80K for the optimally doped YBCO.[9] Gordeev *et al.*[8] suggested that the vortex core diameter $2\xi_c$ becomes smaller than the separation of CuO planes below $T_{cf}$, leading to confinement of vortices between them. In this scenario the vortex liquid freezes below $T_{cf}$ in a smectic phase in which the vortex confinement suppresses thermal fluctuation perpendicular to the CuO layers, while exerting no restraint within the layers, thus inducing crystalline order along the *c*-axis, while remaining liquid within the plane. Despite the strong *ab* peak in $\Delta B_{Irr}(\theta)$ in our sample that appears at about the same temperature observed by Baily *et al.*, we have to take into account the fact that in Ref. 9 a thin film with d=0.2λ(77K) and $\gamma$=6 was investigated. Our disordered and thicker REBCO samples appear to have lower anisotropy ~3 that, if related to a suppressed mass anisotropy, should weaken the intrinsic pinning by the *ab* planes. Yet the observed sudden increase of $\Delta B_{Irr}(\theta)$ at low temperature may suggest matching of the vortex core diameter $2\xi_c$ to the thickness of planar pinning defects in the sample. Indeed, stacking faults would be the most likely source of such narrow *ab* peaks in $\Delta B_{Irr}(\theta)$ because the disruption of the natural sequence of the atomic planes along the *c*-axis and the short $\xi_c$ make the stacking faults much stronger pins than the *ab* planes.

To further address the different nature of the narrow and the wide *ab* peaks in $\Delta B_{Irr}(\theta)$ shown in Fig. 5, we also studied the behavior of resistivity $\rho(B)$ near the depinning transition in which $\rho(B)$ is usually described by the equation $\rho \propto \left(B - B_{Irr}\right)^{s}$, where the exponent *s* determines the shape of the resistive transition. Because in this equation a small change of $B_{Irr}$ significantly affects the estimation of *s*, $B_{Irr}$ was extracted independently from the intercept of the linear fit of $\rho\left[\frac{d\rho}{dB}\right]^{-1}$, as suggested in Ref.15, and the *ρ-B* curves were fitted by the power law equation with $B_{Irr}$ fixed. As shown in Fig. 8, the mean value of *s* is about 3.3±0.2 over almost the whole angular range, except for a small region close to B//*ab*, where *s* decreases to 1.4. Because such reduction of *s* is related to core pinning by planar defects, its effectiveness is inherently limited to the small angular region. When the angle between *B* and *ab* exceeds a few degrees, the vortices form kinked structures which are less strongly pinned than straight vortices parallel to the planar pinning defects.

For the field oriented away from the *ab* planes $s \approx 3.3$ is substantially lower than the 5.3 value found in a proton-irradiated single crystal,[15] where a vortex glass phase is induced by isotropic random point defects. However, $s$=3.3 is similar to the results obtained by Baily *et al.*[9] and Miura *et al.*[10] for strong pinning REBCO films. Moreover, Fig. 8 also does not show any significant change in *s* close to B//*c*, where $B_{Irr}$ is affected by the BZO nanorods.

Now we turn to the *c*-axis peak in $\Delta B_{Irr}(\theta)$ shown in Figs. 4(b) and 6. Here the splayed BZO nanorods result in a broad *c*-axis peak in $\Delta B_{Irr}(\theta)$ and apparently flatten the angular dependence of $B_{Irr}$ over a wide region around the *c*-axis. This flattening is more evident at high temperature so that the magnitude of the *c*-axis peak normalized to the RP contribution $\Delta B_{Irr}(\theta)/B_{Irr,\mathrm{RP}}(\theta)$ decreases as temperature decreases, as shown in Fig. 6. Moreover, the peak in $B_{Irr}$ is not at 180° but at about 190°, which is correlated to the splayed distribution of nanorod orientations shown in Fig. 1. The decreasing relative contribution of the *c*-axis peak at low temperature is at a first sight counterintuitive, because thermal wandering of vortices diminishes upon decreasing temperature and pinning by nanorods becomes stronger.[16] Yet the increase of the ratio $\Delta B_{Irr}(\theta)/B_{Irr,\mathrm{RP}}(\theta)$ at higher temperatures clearly indicates that thermal fluctuations suppress the random component $B_{Irr,RP}(\theta)$ more strongly than the correlated contribution $\Delta B_{Irr}(\theta)$. This effect can be explained by the influence of splayed nanorods which suppress thermal wandering of vortices by blocking proliferation of vortex kinks along the nanorods, thus reducing the effect of thermal fluctuations on $\Delta B_{Irr}$ as compared to the stronger suppression of $B_{Irr,RP}$. This scenario is based on the model proposed by Hwa *et al.*[17] which has been used successfully to explain the strong reduction of flux creep and enhancement of $B_{Irr}$ by splayed columnar defects produced by heavy ion irradiation of the cuprate single crystals.[18] Extensive characterizations of the pinning of BZO-containing and BZO-free variants of these same samples have recently been made.[19,20] Consistent with what found in the present paper, it has been concluded that the correlated pinning due to the BZO nanorods that strongly affect REBCO properties at high temperature becomes weak at low temperature. In fact it has been shown that the angular dependence of $J_c$(4.2K) exhibits no sign of the *c*-axis peak and the RP contribution dominates. Because the thermal fluctuations are negligible at low temperature, the nanorods may act as effective pins also with a strongly misaligned field and they behave in a more isotropic way.

## IV. CONCLUSIONS

We reported systematic studies of the anisotropic irreversibility fields of high-performance thin films with strong pinning nanostructures in very high fields up 45 T. By deconvoluting the RP from the correlated pinning contributions, we found that the former exhibits an effective anisotropy parameter $\gamma_{RP}$~3 rather than $\gamma = 5$-6 characteristic of the mass anisotropy in pure YBCO.[9,10] Moreover, $B_{Irr}(\theta)$ is significantly affected by correlated pinning, which shows interesting temperature dependences, for field both along the *ab* and *c* directions. Correlated pinning for fields aligned close to the *ab* plane is able to nearly double the irreversibility field ($\Delta B_{Irr}$~25T at 78K). At high temperatures the peak in $B_{Irr}$ is determined by magnetic pinning and is slightly affected by a tilt bias in the stacking of *ab*-planar $RE_2O_3$ precipitates,[11,14] whereas below 80 K a significant contribution comes from the vortex core pinning by the stacking faults. The analysis of critical exponents at the depinning transition shows that this last mechanism is effective only in a very narrow region of angles and it becomes negligible once the field is tilted more than 2° away from the *ab* planes. By contrast, at the other field configurations *s* is 3.3 despite the presence of the BZO nanorods near the *c*-axis. This finding suggests that at 78K the contribution to $B_{Irr}$ of correlated pinning along the BZO nanorods is already significantly reduced in favor of a more isotropic RP. For field along the *c*-axis, $B_{Irr}$ is enhanced by ~35% at high temperature but this enhancement loses effectiveness rapidly at lower temperatures, decreasing to ~14% at 56K. Moreover, $J_c$ measurements on similar samples have shown that the nanorods become less effective at 4.2 K both at high and low fields.[19,20] This behavior suggests that nanorods act as correlated pins at high temperature but behave more like random pins at lower temperatures. This phenomenon indicates that the splay distribution of BZO nanorods effectively enhances the correlated pinning contribution to $B_{Irr}$ by suppressing thermal wandering of vortices close to $T_c$.

A portion of this work was performed at the NHMFL, which is supported by NSF Cooperative Agreement No. DMR-0654118, by the State of Florida, and by the DOE. This work was also supported by the U.S. Department of Energy Office of Electricity Delivery and Energy Reliability, Advanced Cables and Conductors.

**APPENDIX**

The angular dependence of the global irreversibility field $B_{Irr}$ ($\theta$) can be calculated for a narrow step-wise distribution F($\phi$) = $1/2\theta_0$ for $|\phi - \theta| < \theta_0$ and F($\phi$) = 0 for $|\phi - \theta| > \theta_0$ where $\theta_0$ quantifies the spread of the local *c*-axis orientations around the angle $\theta$ between **B** and the mean orientation of the *c*-axis. For $\theta_0 << 1$ but $\gamma\theta_0 > 1$, the distribution of the c-axis misorientations can affect the angular dependence of $B_{Irr}$. In this case cos$\phi$ and sin$\phi$ in Eq. (3) can be expanded around the point $\phi = \theta$ up to quadratic terms in u = $\phi - \theta$, which gives after integration:

$$B_{Irr} = \frac{B_{\max}}{2\theta_0 |(\gamma^2 - 1)\cos 2\theta|^{1/2}} \left[ \sin^{-1} \frac{\sin 2\theta + 2\theta_0 \cos 2\theta}{|1 + \cos^2 2\theta + 2\alpha \cos 2\theta|^{1/2}} - \sin^{-1} \frac{\sin 2\theta - 2\theta_0 \cos 2\theta}{|1 + \cos^2 2\theta + 2\alpha \cos 2\theta|^{1/2}} \right], \quad 0 \le \theta \le \pi/4$$

(A1)

$$B_{Irr} = \frac{B_{\max}}{2\theta_0 |(\gamma^2 - 1)\cos 2\theta|^{1/2}} \left[ \sinh^{-1} \frac{\sin 2\theta - 2\theta_0 \cos 2\theta}{|1 + \cos^2 2\theta + 2\alpha \cos 2\theta|^{1/2}} - \sinh^{-1} \frac{\sin 2\theta + 2\theta_0 \cos 2\theta}{|1 + \cos^2 2\theta + 2\alpha \cos 2\theta|^{1/2}} \right], \quad \pi/4 \le \theta \le \pi/2$$

where $\alpha = (\gamma^2 + 1)/(\gamma^2 - 1)$. The effective anisotropy factor $\gamma_{eff} = B_{Irr}(\pi/2)/ B_{Irr}(0)$ obtained from Eq. (A1) results in

$$\gamma_{eff} = \frac{\sinh^{-1}(\theta_0 \sqrt{\gamma^2 - 1})}{\sin^{-1}[(\theta_0 \sqrt{\gamma^2 - 1})/\gamma]} \qquad \text{(A2)}$$

And $\gamma_{eff}$ as a function of $\theta_0$ is shown in Fig. 7.

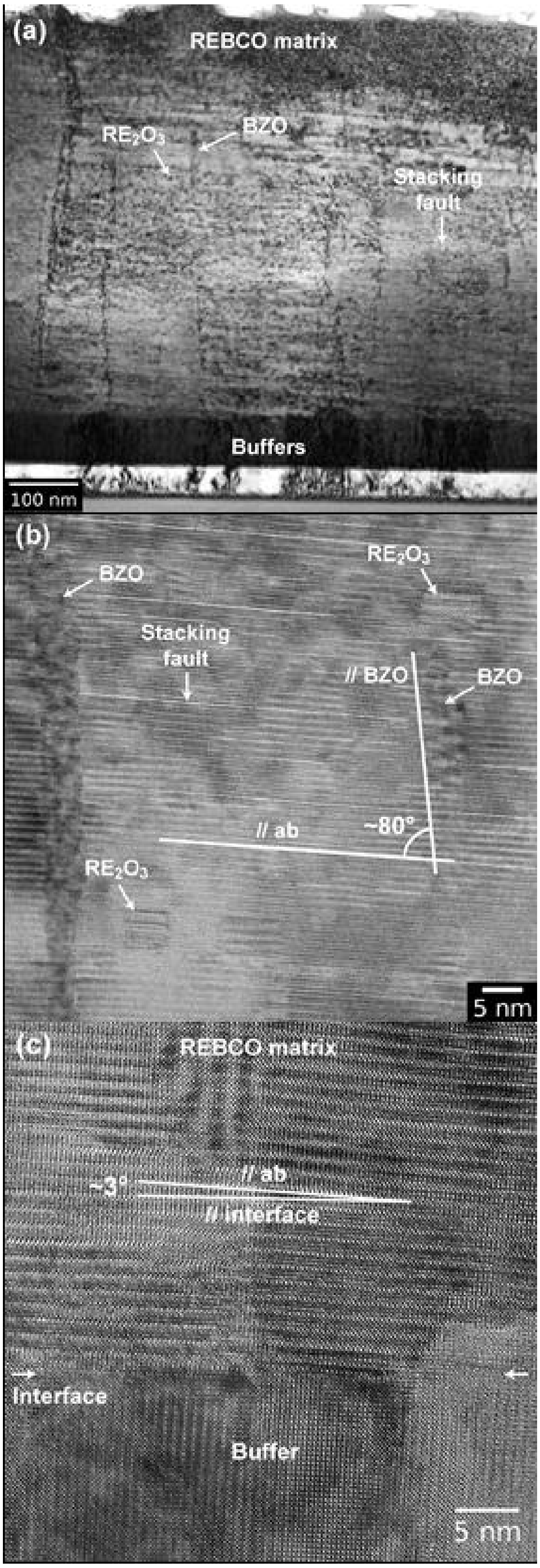

**Fig. 1** Transverse cross-sectional TEM images of the 5%Zr-REBCO sample showing (a) three types of pinning defects: BZO nanorods, $RE_2O_3$ nano-precipitates and stacking faults, (b) the average angle of ~80° between the direction of nanorods and the REBCO *ab*-plane, (c) the REBCO *ab* planes tilted ~3° from the buffer interface.

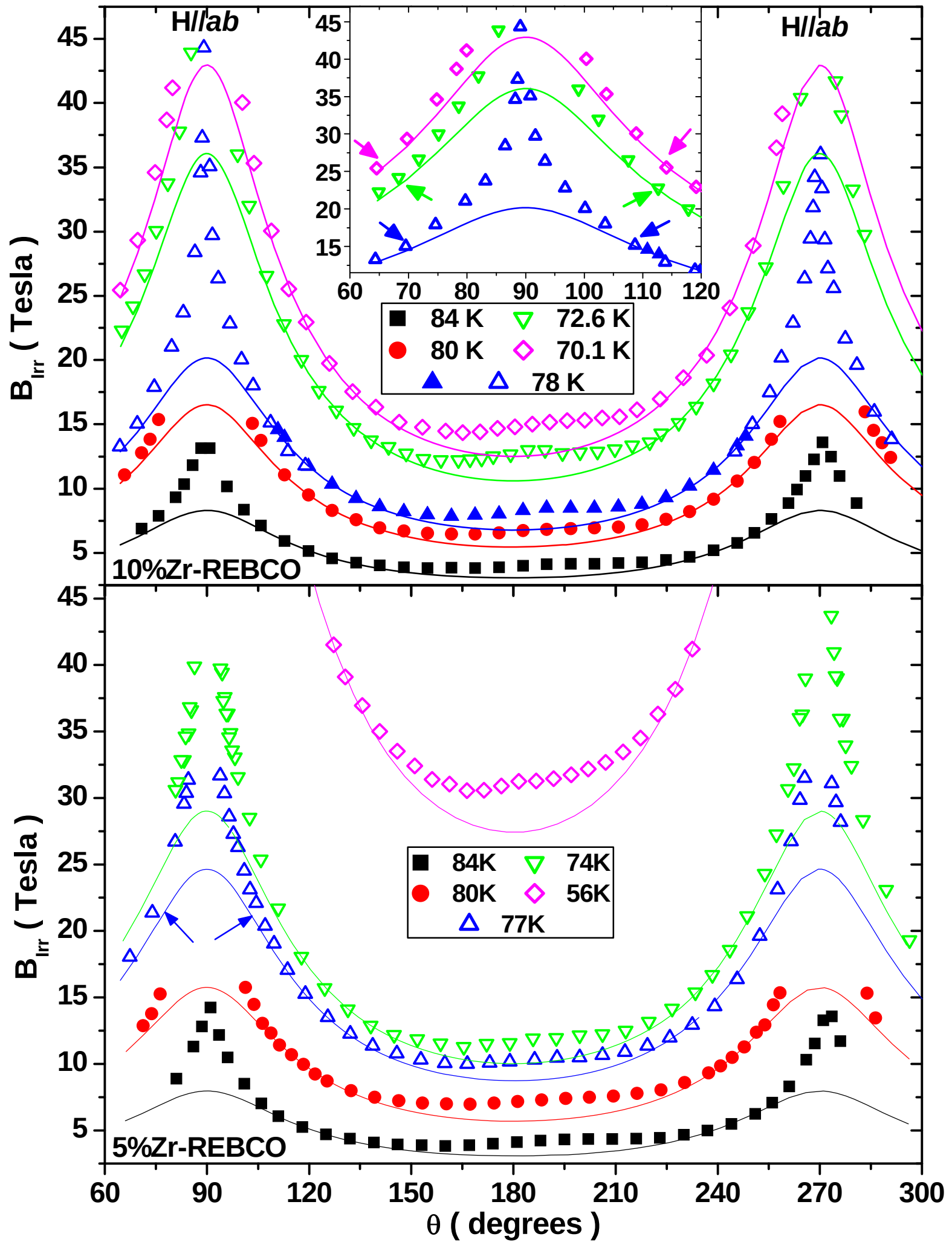

**Fig. 2** Angular dependences of the irreversibility field $B_{Irr}$ (symbols) for a 10%Zr and 5%Zr-REBCO samples and random pinning contribution $B_{Irr,RP}$ to $B_{Irr}$ (lines), determined as explained in the text. In the inset, magnification close to 90° for the 10%Zr-REBCO.

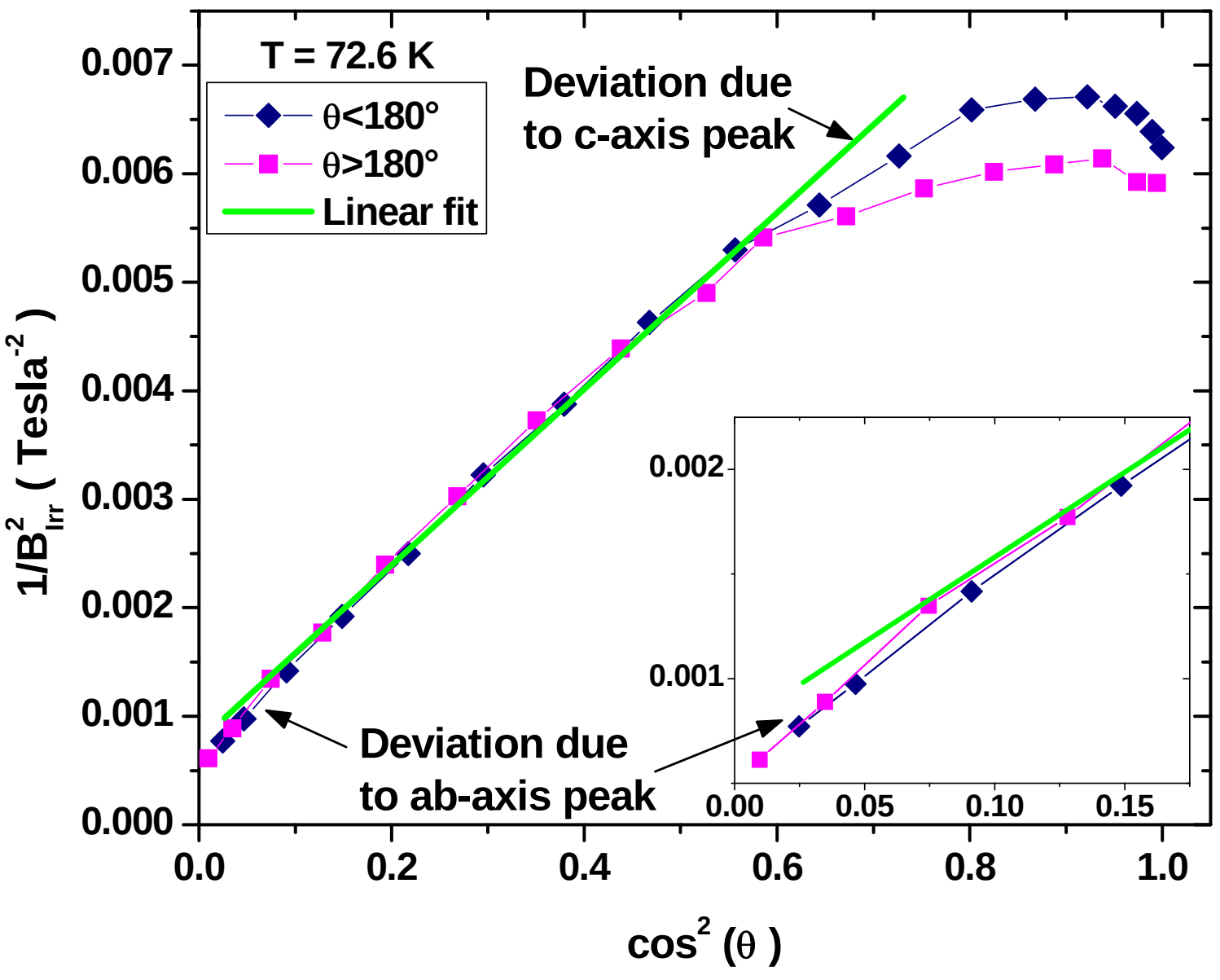

**Fig. 3** $B_{Irr}$ data from Fig. 2 at 72.6 K, replotted as $1/\,B_{Irr}{}^2$ versus $\cos^2(\theta)$ and magnification close to zero in the inset.

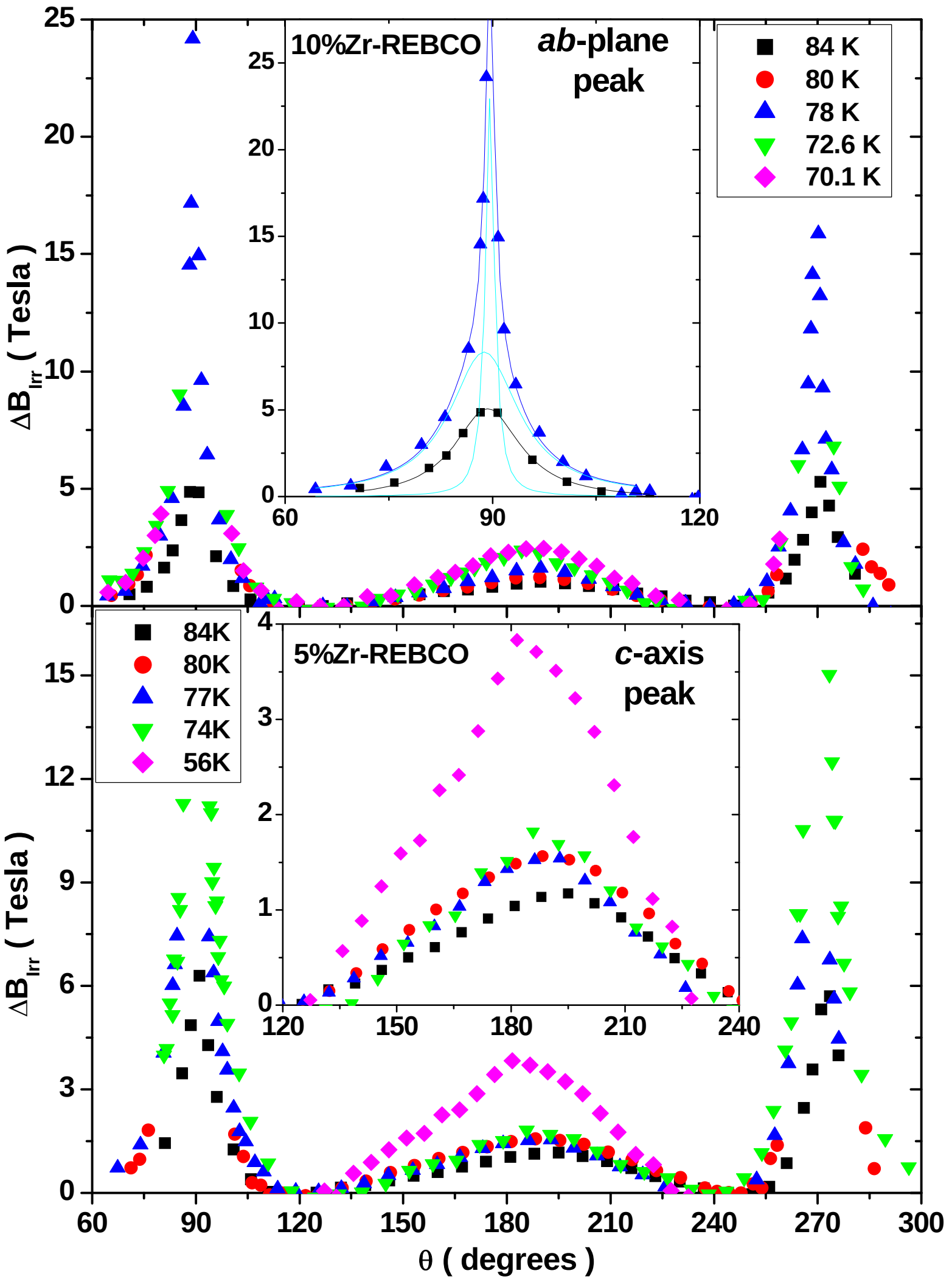

**Fig. 4** Correlated pinning contributions $\Delta B_{Irr}$, determined by subtracting the RP contribution from the experimental data of Fig.2. Inset of (a) shows the single and double peak fits at 84 and 78K, respectively, of the correlated pinning along *ab* for the 10%Zr-REBCO sample. Inset of (b) shows an expanded view of the *c*-axis peak for the 5%Zr-REBCO sample.

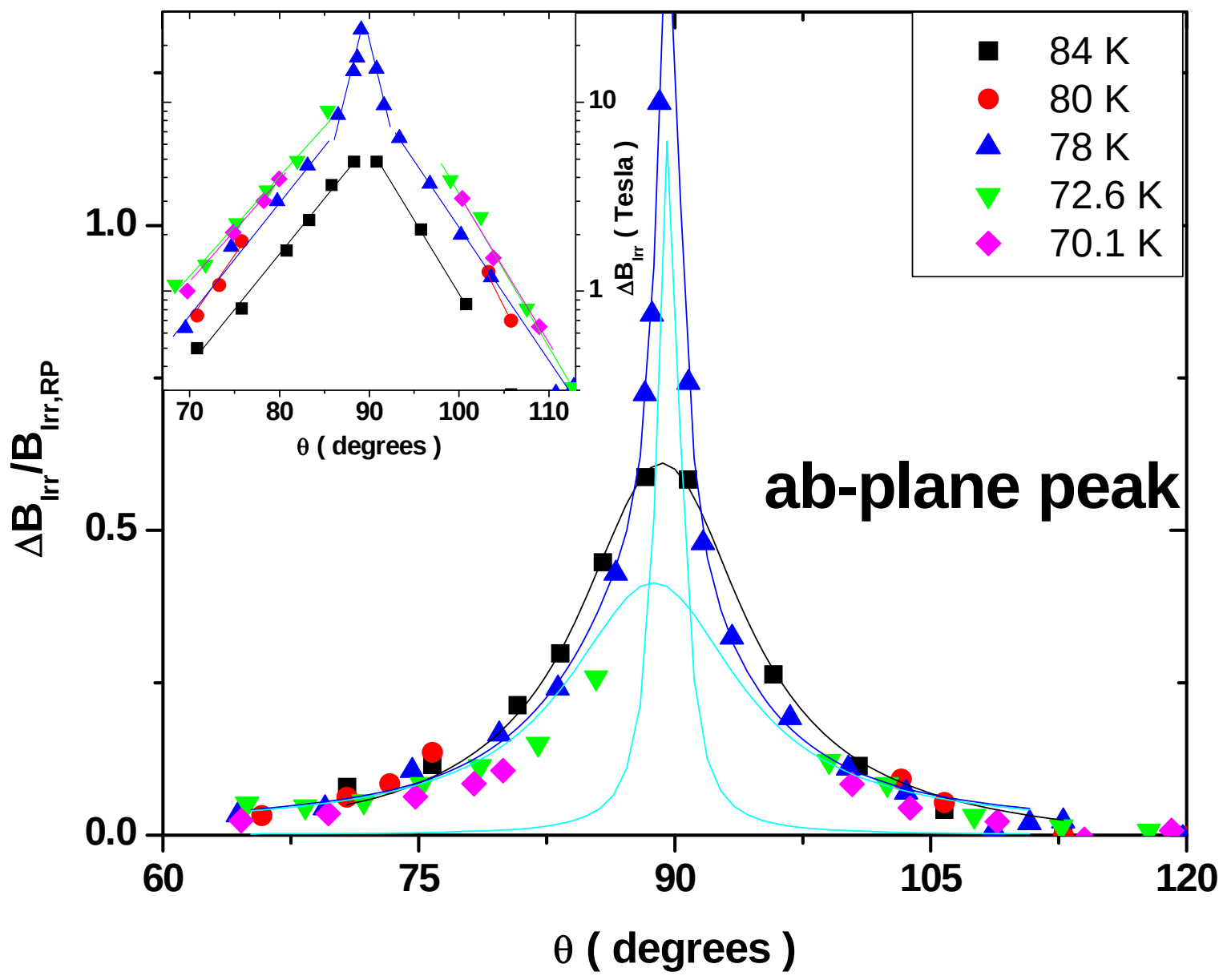

**Fig. 5** Correlated pinning contributions along the *ab* plane plotted on a logarithmic scale, showing the dependence of single- and double-peak structures on temperature.

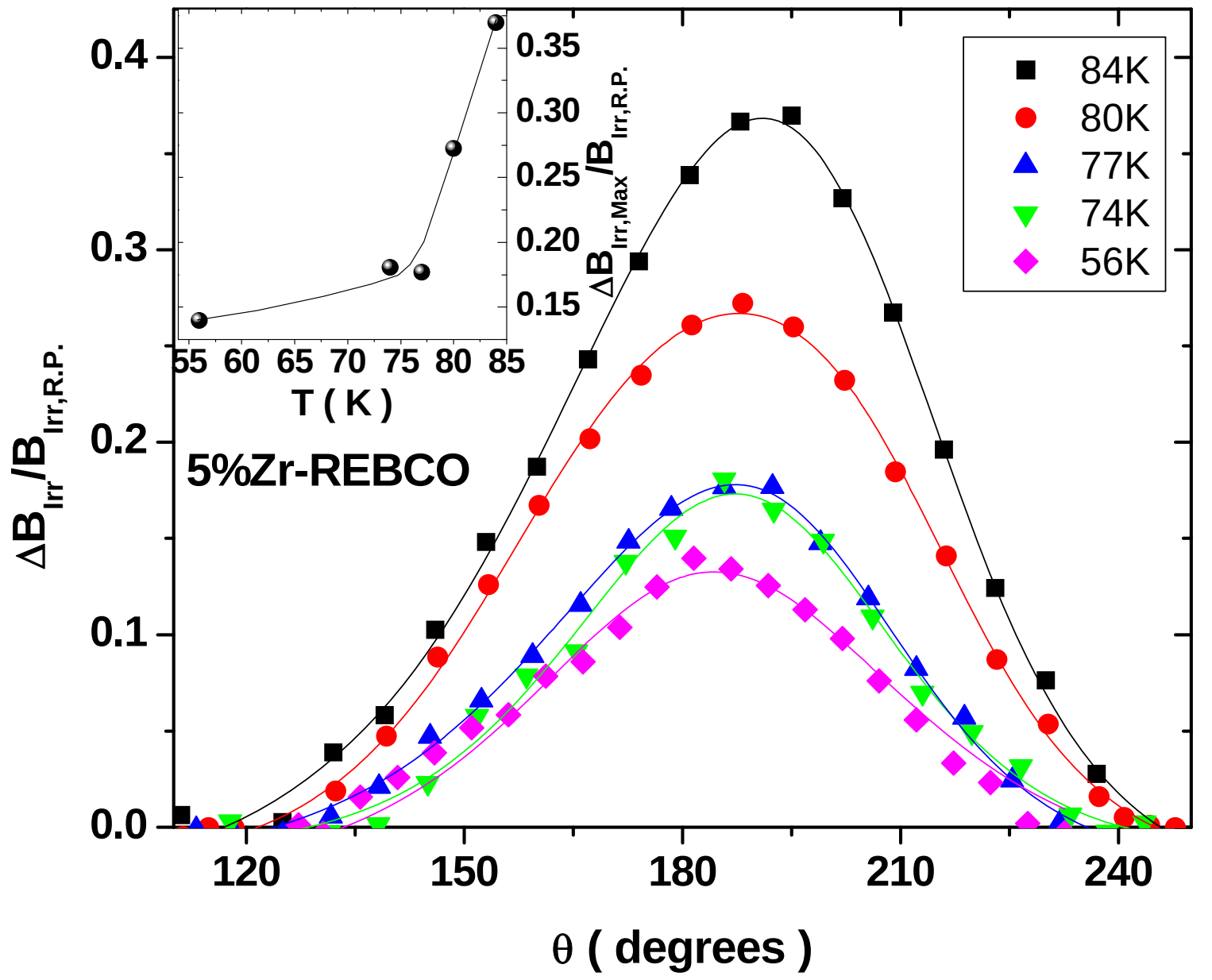

**Fig. 6** Correlated pinning contributions normalized to the RP one close to the *c*-axis; the inset shows the temperature dependence of the peak maxima.

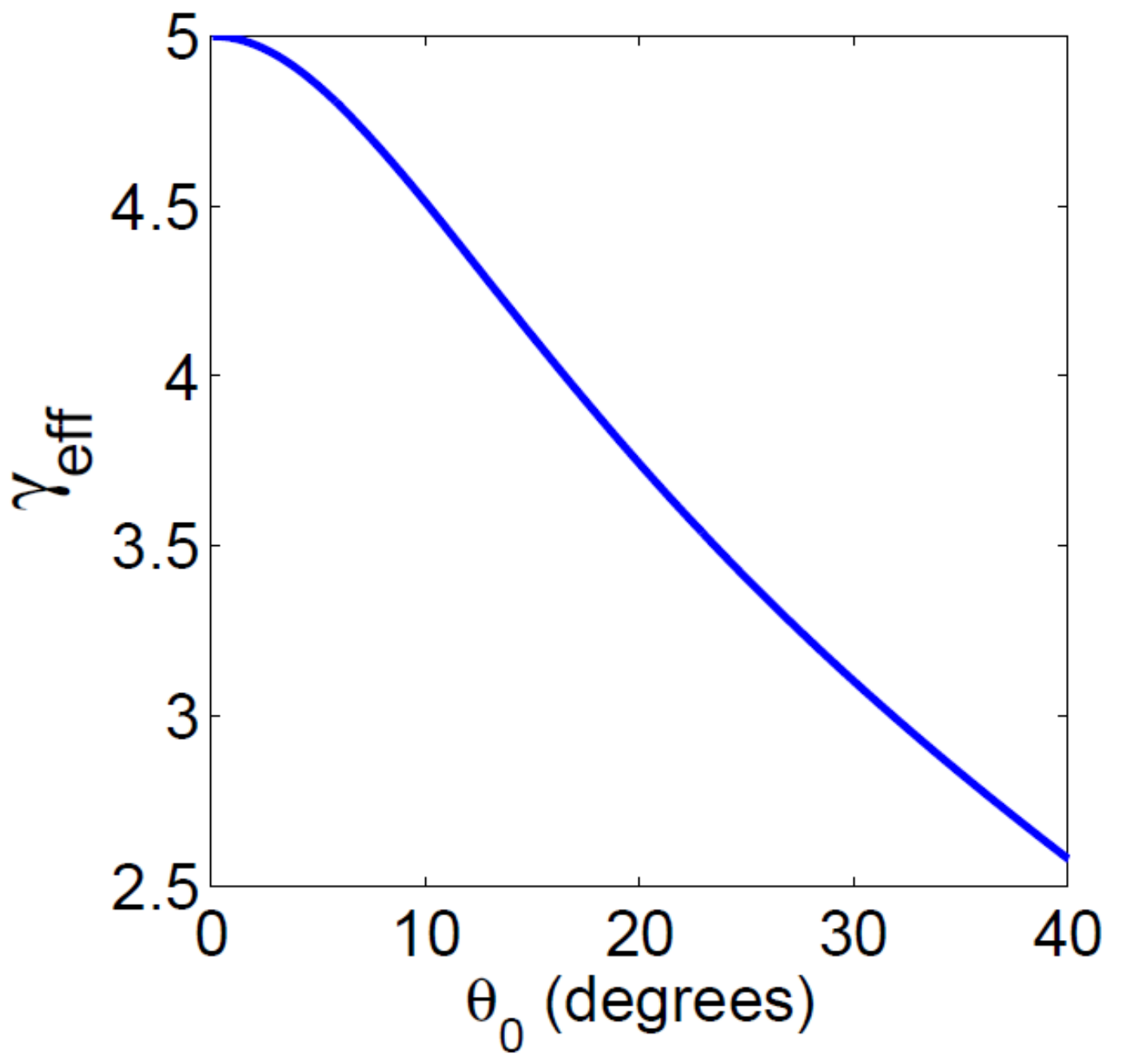

**Fig. 7** Dependence of the effective anisotropy parameter $\gamma_{\rm eff}$ on the width of the c-axis distribution $\theta_0$ defined by Eq. (4) for $\gamma = 5$.

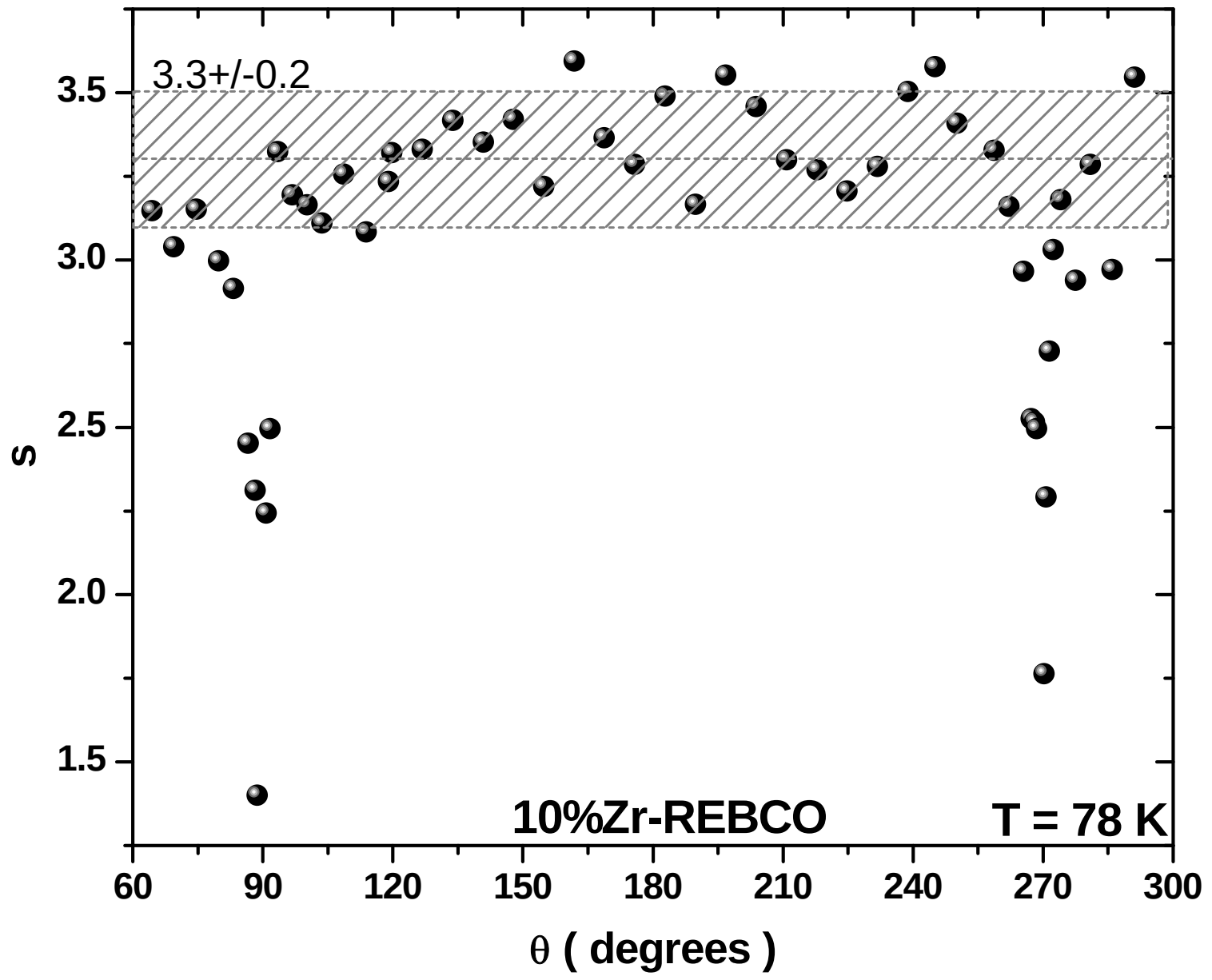

**Fig. 8** Critical exponent *s,* determined from the transition fits with $\rho \propto \left(B - B_{Irr}\right)^{s}$, as a function of angle. The shaded area represents the spread around the mean value over the entire range.